\documentclass[aps,pre,preprint,floatfix,twoside,tightenlines]{revtex4} 
\usepackage{amsmath,graphicx}

\DeclareMathOperator{\Tr}{Tr}
\newcommand{\order}[1]{\mathcal{O}(#1)}
\newcommand{\floor}[1]{\lfloor#1\rfloor}
\newcommand{\deriv}[3][]{\frac{\partial^{#1}#2}{\partial#3^{#1}}}
\newcommand{\fderiv}[3][]{\frac{\mathrm{d}^{#1}#2}{\mathrm{d}#3^{#1}}}

\begin{document}

\title{Quantum free energy differences from non-equilibrium
  path integrals: I.~Methods and numerical application}

\author{Ramses van Zon$^{\ddagger}$, Lisandro Hern\'{a}ndez de la
Pe\~{n}a$^{*\dagger}$, Gilles H. Peslherbe$^\dagger$, and Jeremy
Schofield$^\ddagger$}

\affiliation{$^{\ddagger}$Chemical Physics Theory Group, Department of
Chemistry, University of Toronto, 80 Saint George Street, Toronto,
Ontario M5S 3H6, Canada}

\affiliation{$^*$Department of Chemistry, University of Illinois at
Urbana-Champaign, Urbana, Illinois 61801}

\affiliation{$^{\dagger}$Center for Research in Molecular Modeling and
Department of Chemistry and Biochemistry, Concordia University,
Montreal, Canada}

\date{August 29, 2008}

\begin{abstract}
In this paper, the imaginary-time path integral representation of the
canonical partition function of a quantum system and non-equilibrium
work fluctuation relations are combined to yield methods for computing
free energy differences in quantum systems using non-equilibrium
processes.  The path integral representation is isomorphic to the
configurational partition function of a classical field theory, to
which a natural but fictitious Hamiltonian dynamics is associated.  It
is shown that if this system is prepared in an equilibrium state,
after which a control parameter in the fictitious Hamiltonian is
changed in a finite time, then formally the Jarzynski non-equilibrium
work relation and the Crooks fluctuation relation are shown to hold,
where work is defined as the change in the energy as given by the
fictitious Hamiltonian.  Since the energy diverges for the classical
field theory in canonical equilibrium, two regularization methods are
introduced which limit the number of degrees of freedom to be finite.
The numerical applicability of the methods is demonstrated for a
quartic double-well potential with varying asymmetry.  A general
parameter-free smoothing procedure for the work distribution functions
is useful in this context.
\end{abstract}

\maketitle

\section{Introduction} 

A number of relations valid in the far from equilibrium regime have
appeared in the last fifteen years\cite{Evansetal93,Evansetal94,
GallavottiCohen95a,GallavottiCohen95b,Jarzynski1,Jarzynski2,
Crooks1,Crooks2} that show intriguing relationships between
fluctuations in non-equilibrium systems governed either by
deterministic or stochastic dynamics. Among these relations, the
Jarzynski\cite{Jarzynski1, Jarzynski2} and Crooks
relations\cite{Crooks1,Crooks2} provide a means to compute the free
energy difference between two classical systems by use of a control
parameter that switches the system from one ensemble to another in a
well-defined manner.  The extension of these relations to quantum
systems has been analyzed recently by several
authors\cite{Mukamel1,Maes,Mukamel2, Talkneretal07a,
TalknerHanggi07,DeffnerLutz08}.  Current methods of calculating free
energy differences in quantum systems using non-equilibrium processes
rely on the knowledge of the quantum history of the system which,
while yielding a conceptually appealing picture, cannot provide a
reasonable scheme for the computation of free energy differences in
practical applications.  The challenges associated with constructing
the correct coherent quantum dynamics make the approach difficult to
implement.

The path integral representation of the canonical partition function
is based on mapping a quantum system at finite temperature onto a
classical system with additional degrees of
freedom\cite{Feynman1,Feynman2,BerneThirumalai86}. A non-equilibrium
process can be carried out on this isomorphic classical system along a
well defined trajectory in fictitious time.  As will be demonstrated,
the Jarzynski and Crooks relations are valid for such a process
provided conditions analogous to the ones required for the relations
in a classical system are met.  As a consequence, quantum free
energies using fictitious non-equilibrium classical processes can be
obtained using the path integral representation.  The path integral
formulation, however, involves an infinite number of degrees of
freedom, and, as a result, non-equilibrium dynamical processes in this
representation could lead to divergences which need to be regularized.
The practicality of the method depends sensitively on the rate of
convergence of properties determined from the regularized path
integral to their true quantum values, an issue that is given
considerable attention in this paper and the following paper in the
series, which treats the case of a harmonic oscillator in
detail\cite{subsequent}.

The paper is organized as follows. In Sec.~\ref{system}, the path
integral formalism and two equilibrium methods to compute free energy
differences are reviewed.  The Jarzynski non-equilibrium work relation
and the Crooks fluctuation relation using fictitious non-equilibrium
processes are derived in Sec.~\ref{non-equilibrium}.  In
Sec.~\ref{regularization}, the subtleties associated with the
divergence of the total energy and work of a system with an infinite
number of degrees of freedom are discussed, and two regularization
methods are introduced.  The first one is based on the Fourier
representation of closed paths representing quantum particles and the
second is based on a spatial discretization of the closed paths.  In
Sec.~\ref{numerical-quartic}, the non-equilibrium free energy method
is applied numerically to the free energy of a double well potential
of quartic form as the asymmetry between the wells is varied.  The
conclusions are given in Sec.~\ref{conclusions}.

\section{System and definitions}
\label{system}

Consider a quantum system with a Hamiltonian of the form
$\hat{H}(\lambda)=\hat{T}+\hat{V}$ with $\hat T=\hat p^2/(2m)$ and
$\hat V=V(\hat x,\lambda)$. For simplicity of presentation, the
position operator $\hat x$ and the associated momentum operator $\hat
p$ here are taken to be one-dimensional though the extension of the
analysis to higher dimensional systems is straightforward.  Note that
the potential energy $V$ depends on a control parameter $\lambda$
which is independent of the configuration of the system. The canonical
partition function of this system at an inverse temperature $\beta$ is
defined by
\begin{equation}
   Z(\lambda) = \Tr e^{-\beta \hat{H}(\lambda)},
\label{Zdef}
\end{equation}
which is related to the free energy by
\begin{equation}
   Z(\lambda) = e^{-\beta F(\lambda)}.
\label{ZF}
\end{equation}

The partition function can be written in path integral form
as\cite{Feynman1,Feynman2}
\begin{equation}
   Z(\lambda) = \int\! \mathcal{D}x\:
                e^{-\frac{1}{\hbar}S[x,\lambda]},
\label{partitionFunction}
\end{equation}
where the integral is over closed paths $x(s)$ [i.e.,
$x(\beta\hbar)=x(0)$] and the Euclidean action $S$ is a functional of
$x$ given by
\begin{eqnarray}
   S[x,\lambda] = \int_0^{\beta\hbar}\! \mathrm ds  \left[ \frac12m
                  \Big(\fderiv{x}{s}\Big)^2 + V(x,\lambda) \right].
\label{euclideanAction}
\end{eqnarray}
Here and below the $s$ dependence of $x$ in integrals over $s$ will
always be implied.  Due to the cyclic property of the trace, the
quantum mechanical equilibrium ensemble average $\langle A
\rangle_\lambda^\text{qm}=\Tr\{\hat{A}\exp[-\beta \hat{H}
(\lambda)]\}/Z(\lambda)$ of an operator $A(\hat{x})$ can be written in
the path integral formulation as
\begin{equation}
   \langle A \rangle_{\lambda}^\text{qm} 
   = \frac{1}{Z(\lambda)} \int\! \mathcal{D}x \:
        \overline{A}[x,\lambda]~e^{-\frac{1}{\hbar}S[x,\lambda]}
   = \langle\overline{A}[x,\lambda]\rangle_\lambda
,
\label{average}
\end{equation}
where $\langle \ldots \rangle_\lambda$ denotes a path integral average
$\frac{1}{Z(\lambda)} \int\! \mathcal{D}x \ldots
e^{-\frac{1}{\hbar}S[x,\lambda]}$ and $\overline{A}$ denotes the
imaginary time average
\begin{equation}
  \overline{A}[x,\lambda] 
  = \frac{1}{\beta\hbar}\int_0^{\beta\hbar} \mathrm ds\, A(x(s),\lambda).  
\end{equation}
Note that the path integral average of a function of a single
imaginary time point, $\langle A(x(s^*))\rangle_\lambda$, does not
depend on the choice of the time point $s^*$, due to the imaginary
time translational invariance of the Euclidean action. One may
therefore replace $\langle A(x(s^*))\rangle_\lambda$ by the imaginary
time average $\langle \overline{A}[x]\rangle_\lambda$, which is often
advantageous for reasons of computational efficiency.

Eq.~\eqref{average} defines the equilibrium ensemble average of a
general functional $A[x,\lambda]$ of the path $x(s)$, which could also
be in general a function of $\lambda$.  Note that for
multiple-particle systems, one needs to incorporate exchange effects
associated with quantum statistics\cite{Feynman1,Feynman2}.

In the context of equilibrium statistical mechanics, the free energy
difference between two systems characterized by different values of
the control parameter $\lambda$ can be computed in this picture
through the so-called thermodynamic integration method, and is given
by
\begin{eqnarray}
\Delta F &=& \int_{\lambda_A}^{\lambda_B} d\lambda
\left\langle \frac{1}{\beta\hbar}\int_0^{\beta\hbar}
\mathrm ds \deriv{V(x,\lambda)}{\lambda} \right\rangle_{\lambda}
\nonumber \\
 &=& \int_{\lambda_A}^{\lambda_B} d\lambda 
 \left\langle\: \overline{\deriv{V}{\lambda}}[x,\lambda] \right\rangle_{\lambda}
\label{thermIntegration} ,
\end{eqnarray}
where the free energy change refers to the difference between the
initial and final states through $\lambda$ (i.e. $\Delta F =
F(\lambda_B)-F(\lambda_A)$).  Alternatively, one can compute the free
energy change directly using the following identities\cite{Feynman1}
\begin{eqnarray}
\Delta F &=& -\frac{1}{\beta} \log \left\langle \exp\left[
-\frac{1}{\hbar}(S[x,\lambda_B]-S[x,\lambda_A])\right] \right\rangle_{\lambda_A} 
\nonumber \\
&=& -\frac{1}{\beta} \log \left\langle \exp\left[ -\beta 
\left( \: \overline{V}[x,\lambda_B]-\overline{V}[x,\lambda_A] \right)  \right] \right\rangle_{\lambda_A}
\label{freeEnergyDirect} .
\end{eqnarray}
In the classical limit, the closed paths $x(s)$ transform into a point
$x$ and Eqs.(\ref{thermIntegration}) and (\ref{freeEnergyDirect})
transform into the well-known classical
expressions\cite{Kirkwood35,Zwanzig54}.

\section{Non-equilibrium relations}
\label{non-equilibrium}

The non-equilibrium relations derived in this section pertain to a
non-equilibrium process in fictitious time.  To construct the
fictitious dynamics, a new field $p(s)$ is introduced which is also
periodic in imaginary time, satisfying $p(s)=p(s+\beta\hbar)$.  By
multiplying the path integral representation of the partition function
in Eq.~\eqref{partitionFunction} by
\begin{equation}
\int\! \mathcal{D}p\: e^{-\frac1\hbar\int_0^{\beta\hbar} \mathrm ds\:
               \frac{p^2}{2\mu}} = \frac{1}{C} ,
\label{pdef}
\end{equation}
where $C$ is a normalization constant and $\mu$ is an arbitrary
fictitious mass, one obtains
\begin{equation}
Z(\lambda)  = C \int\! \mathcal{D}x\mathcal{D}p\:
                 e^{-\frac{1}{\hbar}S[x,\lambda]-\frac{1}{\hbar}\int\!
                 \mathrm ds\,\frac{p^2}{2\mu}}. 
\label{Zpx}
\end{equation}
Using Eq.~\eqref{euclideanAction}, this equation can be cast in the form of a
classical partition function
\begin{equation}
  Z(\lambda) =  C \int\! \mathcal{D}x\mathcal{D}p\:e^{-\beta
  H[x,p,\lambda]} ,
\label{Zpx2}
\end{equation}
where the fictitious Hamiltonian is given by
\begin{equation}
  H[x,p,\lambda] = 
  \int_0^1\! \mathrm du  \Bigg[ 
    \frac{p^2}{2\mu} +
    \frac12\kappa\Big(\fderiv{x}{u}\Big)^2 
    + V(x,\lambda) \Bigg].
\label{scaledH}
\end{equation}
Here, a scaled imaginary time variable $u=s/(\beta\hbar)$, has been
introduced, while
\begin{equation*}
  \kappa= \frac{m}{\beta^2\hbar^2}.
\end{equation*}
Equations \eqref{Zpx2} and \eqref{scaledH} correspond to the thermal
field theory of a classical closed elastic string of unit length in
one dimension with mass $\mu$ and string tension $\kappa$.

Recently, Sch\"oll-Paschinger and Dellago showed that Jarzynski's
non-equilibrium work relation\cite{Jarzynski1,Jarzynski2} holds for a
wide class of classical deterministic systems with a finite number of
degrees of freedom\cite{Dellago}. It was demonstrated that if the
dynamics of a system for fixed values of $\lambda$ admits an invariant
distribution of the system plus bath equal to the canonical
distribution at an inverse temperature $\beta$ multiplied by a
function dependent on the bath variables only, then the free energy
difference can be found from
\begin{equation}
  e^{-\beta \Delta F}=\langle\exp(-\beta W)\rangle_{\lambda_A},
\label{JE}
\end{equation}
where the average is over repetitions of a non-equilibrium process in
which the system starts from a configuration drawn from the above
mentioned invariant distribution at $\lambda=\lambda_A$ and is driven
out of equilibrium by varying the control parameter $\lambda$ in a
finite amount of time $\tau$ from $\lambda_A$ to $\lambda_B$ via an
arbitrary protocol $\lambda(t)$.  Furthermore, the work $W$ done in
the process in Eq.~\eqref{JE} is given by
\begin{equation*}
  W = \int_0^\tau \mathrm dt\:\dot{\lambda} \deriv{H}{\lambda}.
\end{equation*}
The result of Sch\"oll-Passinger and Dellago holds for a variety of
different deterministic dynamics. Thus, for the purpose of computing
$\Delta F$ from Eq.~\eqref{JE}, any dynamical evolution scheme can be used.

Perhaps the simplest dynamical evolution is generated by Hamiltonian
dynamics.  In the current context, this dynamics is governed by the
fictitious Hamiltonian in Eq.~\eqref{scaledH}, so the work $W$ is also
fictitious. The equations of motion for the fields $x(u,t)$ and
$p(u,t)$ resulting from the fictitious Hamiltonian are
\begin{subequations}
\begin{align}
  \deriv{x}{t} 
  &= \frac{\delta H[x,p,\lambda]}{\delta p(u)}
   = \frac{p}{\mu} 
\label{dxdt}
\\
  \deriv{p}{t} 
  &= -\frac{\delta H[x,p,\lambda]}{\delta x(u)}
   = \kappa\deriv[2]{x}{u}-\deriv{}{x}V(x,\lambda).
\label{dpdt}
\end{align}
\end{subequations}
Equations~\eqref{dxdt} and \eqref{dpdt} are the usual equations of
motion of a single elastic string in an external potential $V$. It
should be stressed that these equations have no relation to the real
time evolution of the original quantum particle.

Because the system is isolated, the fictitious work done on the system
in changing the control parameter $\lambda$ is precisely the
difference between values of the fictitious Hamiltonian at time $\tau$
and at time $0$.  Introducing the convention that quantities without
explicit time arguments are taken at time zero, one can thus write
\begin{equation}
  W = H[x(\tau),p(\tau),\lambda_B]-H[x,p,\lambda_A].
\label{Wdef}
\end{equation}
Now consider the exponential average of $W$ over initial conditions
drawn from the canonical equilibrium of the string at
$\lambda=\lambda_A$:
\begin{align}
  \big<e^{-\beta W}\big>_{\lambda_A}
  &= 
   \frac{
     \int\! \mathcal{D}x\mathcal{D}p\:e^{-\beta W}e^{-\beta H[x,p,\lambda_A]}
   }{
     \int\! \mathcal{D}x\mathcal{D}p\:e^{-\beta H[x,p,\lambda_A]}
   }
\nonumber\\
  &= 
   \frac{
     \int\! \mathcal{D}x\mathcal{D}p\:e^{-\beta H[x(\tau),p(\tau),\lambda_B]}
   }{
     \int\! \mathcal{D}x\mathcal{D}p\:e^{-\beta H[x,p,\lambda_A]}
   }.
\nonumber
\end{align}
In the numerator, one can change path integration variables from the
initial field $x$ and $p$ to $x'=x(\tau)$ and $p'=p(\tau)$. The
Jacobian of this transformation is equal to unity due to Liouville's
theorem\cite{Goldstein}, so that
\begin{align}
  \big<e^{-\beta W}\big>_{\lambda_A}
  &= 
   \frac{
     \int\!\mathcal{D}x'\mathcal{D}p'\:e^{-\beta H[x',p',\lambda_B]}
   }{
     \int\!\mathcal{D}x\mathcal{D}p\:e^{-\beta H[x,p,\lambda_A]}
   }
\nonumber\\
   &= \frac{Z(\lambda_B)}{Z(\lambda_A)} = e^{-\beta\Delta F},
\label{JEH}
\end{align}
which is Jarzynski's non-equilibrium work relation, i.e. Eq.~\eqref{JE}. It
should be stressed that while $W$ is fictitious work, the resulting
$\Delta F$ is the real quantum free energy difference.  Note that for
a process that occurs infinitely fast, i.e.  the switching time
$\tau=0$, one recovers Eq.~\eqref{freeEnergyDirect}. For an infinitely slow
process, however, one doesn't recover the canonical equilibrium
expression Eq.~\eqref{thermIntegration} due to the fact that the system
evolves in isolation.

Another non-equilibrium relation, the Crooks fluctuation
relation\cite{Crooks1,Crooks2}, can also be shown to hold in this
context.  Consider a non-equilibrium process performed as stated
above, as well as in the reversed sense, i.e.\ starting from
configurations drawn from a canonical distribution at
$\lambda=\lambda_B$ and driven out of equilibrium by varying $\lambda$
in time from $\lambda_B$ to $\lambda_A$ in the reversed direction of
time, that is, with $\lambda(t)\to\lambda(\tau-t)$.  Then the
probability that an amount $W$ of work is done during the forward
process can be written as
\begin{align}
  &P_f(W)  
\nonumber\\
  &= 
     \int\! \mathcal{D}x\mathcal{D}p~\frac{e^{-\beta H[x,p,\lambda_A]}}{Z_A}\delta\left(W-\int_0^\tau \mathrm dt\:\dot{\lambda} \deriv{H}{\lambda}\right)
\nonumber\\
  &=
     \int\! \mathcal{D}x'\mathcal{D}p'~\frac{e^{-\beta W}e^{-\beta H[x',p',\lambda_B]}}{Z_A}\delta\left(W+\int_0^\tau \mathrm dt\:\dot{\lambda} \deriv{H}{\lambda}\right)
\nonumber\\
  &= e^{\beta W}e^{-\beta\Delta F} P_r(-W),
\label{CFR}
\end{align}
where $P_r(-W)$ is the probability that an amount of work $-W$ is done
during the reverse process.  Eq.~\eqref{CFR} is known as the Crooks
fluctuation relation\cite{Crooks1,Crooks2}. Note that the value of $W$
at which the two distributions $P_f(W)$ and $P_r(-W)$ become equal,
which will be denoted by $W_c$, is precisely when $W_c=\Delta F$. Thus
this relation allows $\Delta F$ to be computed by determining where
the plots of $P_f(W)$ and $P_r(-W)$ versus $W$ intersect. This
approach is known as the crossing method\cite{Collinetal05,Ritort05}.

The Crooks fluctuation relation, Eq.~\eqref{CFR}, can be extended to a
conditional ensemble in which a particular value of a variable, called
the ``reaction coordinate'', is held fixed to a value
$\chi$\cite{PMF}. The free energy $f(\chi)$ at this constrained value
of the reaction coordinate is known as the potential of mean force.
Here, we consider the reaction coordinate $\chi$ to be given by a
functional $\tilde\chi$ of the position-field $x$, and the Hamiltonian
to take the form
\begin{align}
H[x,p,\lambda]=H_0[x,p]+\Phi(\tilde\chi[x],\lambda). 
\label{HPMF}
\end{align}
The potential of mean force $f(\chi)$ is then defined as
\begin{align}
  e^{-\beta f(\chi)} =\langle\delta(\chi-\tilde\chi[x])\rangle_0
\label{PMF}
\end{align}
plus an arbitrary constant, where the average is over the path
integral corresponding to $H_0$. Defining $P_f(W,\chi)$ as the joint
probability that work $W$ is done in the forward process with a final
value of $\chi$ for the reaction coordinate, and $P_r(W,\chi)$ as the
joint probability for work $W$ and initial value $\chi$ in the reverse
process, then one can derive analogously to Eq.~\eqref{CFR} that
$P_r(-W,\chi)= e^{-\beta W}e^{\beta\Delta F} P_f(W,\chi)$, again
using Liouville's theorem.
Integrating this identity over $W$ gives
\begin{align}
\langle\delta(\chi-\tilde\chi[x])\rangle_{\lambda_B}
=\langle\delta(\chi-\tilde\chi[x(\tau)])e^{-\beta (W-\Delta F)}\rangle_{\lambda_A}
.
\label{tmp1}
\end{align}
Since the objective is to get the potential of mean
force, the appearance of $\Delta F$ seems to be a
complication. However, the left hand side of Eq.~\eqref{tmp1} does not
correspond to the potential of mean force in Eq.~\eqref{PMF}, since
the $H_B$ and $H_0$ ensembles are different. {}Instead, from
Eq.~\eqref{HPMF}, one easily shows that $
\langle\delta(\chi-\tilde\chi[x])\rangle_{\lambda_B} =
\frac{Z_0}{Z_B}e^{-\beta\Phi(\chi,\lambda_B)}
\langle\delta(\chi-\tilde\chi[x])\rangle_0, $ and since
$e^{-\beta\Delta F}=Z_B/Z_A$, Eq.~\eqref{tmp1} becomes
\begin{align}
e^{-\beta f(\chi)}
&=c\left\langle\delta(\chi-\tilde\chi[x(\tau)])e^{-\beta[W-\Phi(\chi,\lambda_B)]}\right\rangle_{\lambda_A}
,
\end{align}
which is the path-integral analog of the result of Paramore et
al.\cite{PMF}. Note that $c=\frac{Z_A}{Z_0}$ is a constant, which does
not matter for the potential of mean force, so that $f(\chi)$ can be
determined from a non-equilibrium process in a similar way as the free
energy $\Delta F$.

\section{Regularization Methods}
\label{regularization}

The invariance of the volume element $\mathcal{D}x \mathcal{D}p$ under
Hamiltonian dynamics is essential to derive the Jarzynski equality in
\eqref{JEH}, the Crooks fluctuation relation in
\eqref{CFR} and its extension to constrained ensembles. 
However, strictly speaking the phase space volume element
is infinite here. The infinite volume of the phase space volume
element is reminiscent of ultraviolet divergences in classical field
theories that arise from an infinite dimensional phase space.  For
instance, the average kinetic energy of a classical elastic string is
equal to the number of degrees of freedom times $1/(2\beta)$, but
since the number of degrees of freedom of the string is infinite, the
average kinetic energy diverges.  The ultraviolet divergences present
difficulties in the direct application of the results of
Sch\"oll-Paschinger and Dellago\cite{Dellago} to the elastic string,
because in the definition of the work \eqref{Wdef}, both
$H[x(\tau),p(\tau),\lambda_B]$ and $H[x(0),p(0),\lambda_A]$ are
divergent quantities, so $W$ might not be well defined. Other
complications would arise for dynamics with phase space contraction,
which are not considered here. 

To assess whether the fluctuation relations derived in the previous
section are meaningful for a system with an infinite number of degrees
of freedom, the divergences need to be regularized such that a finite
number $M$ of degrees of freedom results. There are two general
approaches to regularizing the path integral: the first is a
\emph{Fourier regularization}, in which a wave vector cut-off in
Fourier space is introduced, while the other is a \emph{bead
regularization}, which consists of discretizing the points of the
elastic string by replacing the continuous imaginary time $s$ by a
finely spaced lattice of imaginary time points. In order to establish
a proper theory, the regularized quantities such as the free energy
and the distribution of fictitious work must converge to a finite
limit as the $M\to\infty$, corresponding to the imaginary time lattice
spacing going to zero or the cut-off to infinity, respectively. Such
regularization is also required to obtain feasible computational
methods, and for these the \emph{nature} of the convergence (of the
free energy, the work distribution etc.) to a finite result is
important for the efficiency of the method.

While taking a finite value for $M$ solves the infinite phase space
volume problem for the partition function using either of the two
regularization procedures, it is a separate question whether the
distribution of work values is well defined in the limit
$M\to\infty$. The work distribution is used in the Jarzynski and
Crooks relations, and is central to the non-equilibrium methods.  For
any finite $M$, this distribution will be well defined and yield
information on the finite-$M$ free energy difference $\Delta
F_M$. While $\lim_{M\to\infty}\Delta F_M$ is equal to the true quantum
free energy difference $\Delta F$, it must be noted that there is
currently no general method to show that $P(W)$ is well behaved as
$M\to\infty$.  In Sec.~\ref{numerical-quartic}, it is demonstrated
numerically that the work distribution converges for an asymmetric
double-well potential.  In the companion paper\cite{subsequent}, the
convergence of the work distribution will be shown analytically for
the specific case of a particle in a harmonic well of changing
strength, which can be solved exactly.

\subsection{Fourier regularization}

The central idea of the Fourier space regularization approach is to
restrict the number of Fourier components $\tilde x_k$ and $\tilde
p_k$ of the continuous fields $x(u)$ and $p(u)$ to be finite, where
\begin{subequations}
\begin{align}
\tilde x_k&=(\mathcal F x)_k
\label{deftildexk1}
\\
\tilde p_k&=(\mathcal F p)_k, 
\end{align}
\end{subequations}
and
\begin{equation*}
  (\mathcal F f)_k = \int_0^1\!\mathrm du\: e^{2\pi\mathrm i k u}
  f(u).
\end{equation*}
Note that because $x$ and $p$ are periodic with period $1$, $k$ only
takes integer values. Furthermore, since $x$ and $p$ are real fields,
their Fourier modes satisfy $\tilde x_{-k}=\tilde x^*_k$ and $\tilde
p_{-k}=\tilde p^*_k$. In the Fourier representation, the Hamiltonian
\eqref{scaledH} becomes
\begin{equation}
  H(\tilde{\mathbf x},\tilde{\mathbf p},\lambda)
  = 
  \sum_{k=-\infty}^\infty 
  \Big(
        \frac{|\tilde p_k|^2}{2\mu}
	+
	\frac12m\omega_k^2|\tilde x_k|^2
  \Big)
  + \tilde V(\tilde{\mathbf x},\lambda) ,
\label{Hn}
\end{equation}
where $\tilde{\mathbf x}$ and $\tilde{\mathbf p}$ are the collections
of all $\tilde x_k$ and $\tilde p_k$, respectively, and the
dispersion relation is
\begin{equation}
  \omega_k = 2\pi k\sqrt{\frac{\kappa}{m}} = \frac{2\pi k}{\hbar\beta}
  \label{dispersion-Fourier}
\end{equation}
and 
\begin{equation}
\tilde V(\tilde{\mathbf x},\lambda) = \mathcal F(V(\mathcal
F^{-1}\tilde{\mathbf x},\lambda))_{k=0}. 
\label{tildeVdef}
\end{equation}
Note that for $k\neq0$ modes $k$ and $-k$ are degenerate. 
Given the Taylor series for the potential,
$V(x,\lambda)=a_\lambda+b_\lambda x+c_\lambda
x^2+ d_\lambda x^3+\dots$, from Eq.~\eqref{tildeVdef} one gets
\begin{align}
  \tilde V(\tilde{\mathbf x},\lambda)
  &= a_\lambda+b_\lambda \tilde x_0
+c_\lambda\sum_{k=-\infty}^\infty|\tilde x_k|^2
\nonumber\\
&\quad+d_\lambda
\sum_{k_1=-\infty}^\infty
\sum_{k_2=-\infty}^\infty
\tilde x_{k_1}\tilde x_{k_2}\tilde x^*_{k_1+k_2}
+\ldots
\label{Vtilde}
\end{align}

Because of the periodic boundary conditions imposed on the fields
$x(u)$ and $p(u)$, the infinite volume in phase space is now
countable. We can thus regularize the theory by imposing a cut-off
$k_c$ on the values of $k$. With a finite cut-off, the application of
Liouville's theorem to derive Eqs.~\eqref{JEH} and \eqref{CFR} poses no
problem since the Hamiltonian flow involving $M=1+2k_c$ degrees of
freedom preserves phase-space volume for any finite $M$, and hence
also in the limit $M\to\infty$. Note that the limit $M\to1$
corresponds to the classical limit, as can be seen by taking only the
$k=0$ term in Eq.~\eqref{Hn}.

Using this Fourier space regularization, the free energy converges in
the limit $M\to\infty$ as $\order{M^{-1}}$ when used in the above
straightforward form\cite{Eleftheriouetal99}. Using so-called partial
averaging techniques\cite{Dolletal85}, this can be turned into a
$\order{M^{-2}}$ convergence\cite{Eleftheriouetal99}. However, the
next regularization approach, based on replacing the string by a set
of beads connected by springs, is a much simpler and more general way
to get an $\order{M^{-2}}$ convergence (or higher).

\subsection{Bead regularization}

In contrast with the Fourier regularization, in the bead
regularization procedure, the closed path $x(u)$ is represented by a
lattice of $M$ points. The points $u=n/M$ of the lattice are called
beads, and their positions and momenta are denoted by $\mathbf x
\equiv \{x_1,\dots,x_M\}$ and $\mathbf p=\{ p_1,\dots,p_M\}$,
respectively. Using these variables, the partition function
$Z(\lambda)$ in Eq.~\eqref{Zpx2} can be approximated
by\cite{BerneThirumalai86}
\begin{equation}
Z_M(\lambda) =
\left(\frac{mM}{4\pi^2\hbar^2 \mu'}\right)^{M/2}
 \int\!\mathrm d\mathbf x \mathrm d\mathbf p
 ~e^{-\beta H_M(\mathbf x,\mathbf p,\lambda)},
 \label{discreteZ}
\end{equation}
where
\begin{equation}
 H_M(\mathbf x,\mathbf p,\lambda) =
\sum_{n=1}^M \left[ \frac{p_n^2}{2 \mu'}+\frac12\kappa M 
		    (x_n-x_{n+1})^2
		   +\frac{1}{M} V(x_n,\lambda) \right]
\label{discreteH}
\end{equation}
and $x_{M+1}=x_1$, i.e., the $x_n$ form a ``ring polymer.'' In
Eqs.~\eqref{discreteZ} and \eqref{discreteH}, $\mu'$ is the mass associated
with each bead. To ensure that the ring polymer approaches the elastic
string limit as $M\to\infty$, the mass of each of the $M$ beads has to
be $\mu'=\mu/M$, but for numerical applications at finite $M$, $\mu'$
is a free parameter.

This regularization scheme is the basis of the frequently used path
integral molecular dynamics (PIMD) method\cite{ParrinelloRahman84,
BerneThirumalai86,Tuckermanetal93,Minaryetal03} for computing
canonical equilibrium averages for quantum systems.

To derive the bead regularization of the partition function in
Eq.~\eqref{discreteZ} and the fictitious Hamiltonian in Eq.~\eqref{discreteH}, one
usually starts from the Trotter formula for the Boltzmann operator,
i.e.\cite{Trotter,Suzuki76}
\begin{equation}
e^{-\beta \hat H}=\lim_{M\to\infty}
\left(e^{-\beta \hat V/M}e^{-\beta \hat T/M}\right)^M.
\label{trotter}
\end{equation}
However, it is hard to see from Eq.~\eqref{trotter} why the convergence of the
free energy for large $M$ would behave $\order{M^{-2}}$, as mentioned
above.  The convergence properties are easier to determine starting
from the symmetric version
\begin{equation}
e^{-\beta \hat H/M} = 
e^{-\beta\hat V/(2M)}e^{-\beta\hat T/M}e^{-\beta\hat V/(2M)}
+\order{M^{-3}},
\label{verlet}
\end{equation}
which follows from the Baker-Campbell-Hausdorff formula\cite{Parisi}.
Equation \eqref{verlet} allows the Boltzmann operator to be expressed
as
\begin{equation}
e^{-\beta\hat H} = \left[
e^{-\beta\hat V/(2M)}e^{-\beta\hat T/M}e^{-\beta\hat V/(2M)}\right]^M
+\order{M^{-2}}. 
\label{verlettotal}
\end{equation}
When taking the trace of the Boltzmann operator to obtain the
partition sum in Eq.~\eqref{Zdef}, the first term on the right-hand side of
Eq.~\eqref{verlettotal} can be rewritten using the cyclic properties of the
trace in the form of the Trotter formula \eqref{trotter}. Thus, for
the partition sum, the two so-called splitting methods \eqref{trotter}
and \eqref{verlettotal} lead to the same result. Since the latter
converges as $\order{M^{-2}}$, so does the former. Note that if the
trace is not taken, such as for the imaginary time propagator or for
expectation values of operators which do not commute with $\hat x$,
the two splitting methods exhibit different convergence
behavior\cite{ZhaoPan89}.  In particular, the convergence behavior of
the Trotter form then becomes $\order{M^{-1}}$\cite{Suzuki76}.

Taking the trace to get the partition sum, and the standard technique
of inserting closure
relations\cite{Feynman1,Feynman2,BerneThirumalai86}, one finds
\begin{equation}
  Z_M(\lambda) = \left(\frac{mM}{2\pi\beta\hbar^2}\right)^{M/2}
  \int\!\mathrm d\mathbf x \:e^{-\beta\sum_{n=1}^M
  \left[\frac{mM}{2\hbar^2\beta^2}(x_{n+1}-x_{n})^2+M^{-1}V(x_n)\right]}.
\label{almost}
\end{equation}
Finally, multiplying the right-hand side of Eq.~\eqref{almost} with 
\begin{equation}
  1 = \left(\frac{\beta}{2\pi\mu'}\right)^{M/2}\int\!\mathrm d\mathbf p\:
  e^{-\beta\sum_{n=1}^M \frac{p_n^2}{2\mu'}}
\label{MB}
\end{equation}
one obtains Eqs.~\eqref{discreteZ} and \eqref{discreteH}, and one sees that
$Z_M(\lambda)$ converges to $Z(\lambda)$ as $\order{M^{-2}}$.

The bead regularization in Eq.~\eqref{discreteH} is not as different from the
string's Fourier representation in Eq.~\eqref{Hn} as it may appear at first
sight, as becomes clear when Eq.~\eqref{discreteH} is written in the Fourier
representation as well. Since the beads are discrete, the Fourier
transform is also discrete and takes the form
\begin{subequations}
\begin{align}
  \tilde x_k &= \frac1M\sum_{n=1}^M e^{2\pi\mathrm i kn/M} x_n
\label{deftildexk2}
\\
  \tilde p_k &= \sum_{n=1}^M e^{2\pi\mathrm i kn/M} p_n,
\label{deftildepk2}
\end{align}
\end{subequations}
where $k$ runs from $0$ to $M-1$. Because $\tilde x_k=\tilde x_{M+k}$,
one can also choose the range of $k$ to be from $-\floor{M/2}$ to
$\floor{M/2}$, and this is more convenient since a natural cut-off
$k_c=\floor{M/2}$ then arises.  The asymmetry between the position and
the momentum transformation in Eqs.~\eqref{deftildexk2} and
\eqref{deftildepk2} is necessary to have the Fourier transformation
preserve the canonical structure, while at the same time letting the
definition of $\tilde x_k$ in Eqs.~\eqref{deftildexk1} and
\eqref{deftildexk2} coincide for $M\to\infty$.  Applying this
transformation to the fictitious Hamiltonian in Eq.~\eqref{discreteH} gives
\begin{equation}
  H(\tilde{\mathbf x},\tilde{\mathbf p},\lambda) 
  = 
  \sum_{k=-k_c}^{k_c} 
  \Big(
        \frac{|\tilde p_k|^2}{2M\mu'}
	+
	\frac12m\omega_k^2|\tilde x_k|^2
  \Big)
  + \tilde V(\tilde{\mathbf x},\lambda) ,
\label{Hn2}
\end{equation}
where the only real difference with the fictitious elastic string
Hamiltonian \eqref{Hn} (with a wave vector cut-off $k_c$) lies in the
dispersion relation
\begin{equation}
  \omega_k = \frac{2M}{\hbar\beta}\sin\frac{\pi k}{M}
  \label{dispersion-beads}
\end{equation}
instead of Eq.~\eqref{dispersion-Fourier}. Note that for $k\neq 0$, $k$ and
$-k$ are again degenerate, with the exception that for even $M$, the
mode $k=-M/2$ is identical to the mode $k=M/2$ and only one of these
should be included. Naturally, in the limit $M\rightarrow\infty$,
Eq.~\eqref{dispersion-beads} reduces to Eq.~\eqref{dispersion-Fourier} for
fixed~$k$.

The potential term $\tilde V$ in Eq.~\eqref{Hn2} is given by the same
expression \eqref{Vtilde} as for the elastic string, with all sums
over wave vectors cut-off at $k_c$.  Therefore, both for the Fourier
and the bead regularization, the
equations of motion in terms of $\tilde x_n$ and $\tilde p_n$ take the
form
\begin{subequations}
\begin{align}
  \fderiv{\tilde x_k}{t} 
  =& \frac{\tilde p_k}{\mu''} 
\\
  \fderiv{\tilde p_k}{t} 
  =&
  -\omega_k^2 \tilde x_k - b_\lambda\delta_{k0}-2c_\lambda \tilde x_k
  - 3d_\lambda\sum_{q=-k_c}^{k_c}\tilde x_{q}\tilde x_{k-q}
-\dots
\end{align}
\end{subequations}
where $\mu''=\mu$ for the Fourier regularization and $\mu''=M\mu'$ for
the bead regularization.

\section{Path integral simulations}
\label{numerical-quartic}

\begin{figure}[t]
\includegraphics[width=0.9\columnwidth]{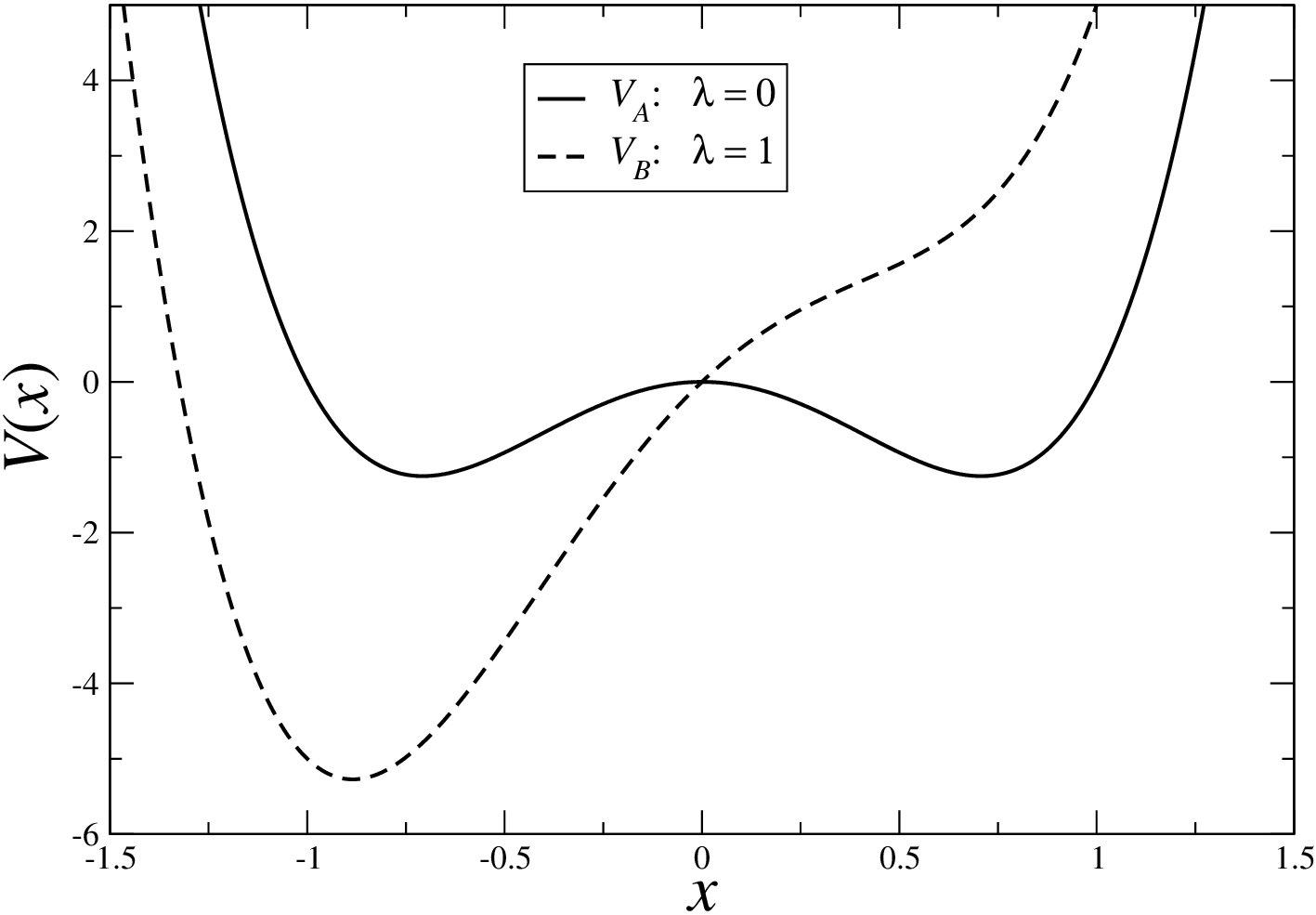}
\caption{Two model potentials confining a one dimensional quantum
         particle, for which the free energy difference is determined
         by non equilibrium methods in Sec.~\ref{numerical-quartic}.
         Note that the $\lambda$ values given correspond to
         $V(x,\lambda = 0) = V_A$ and $V(x,\lambda = 1) = V_B$.}
\label{potentials}
\end{figure}

As an illustration of calculations of free energy differences in
non-trivial quantum systems, consider the difference in free energy
between a quantum particle confined in a symmetric quartic double-well
potential $V_A(x)$ and a quartic potential with a linear bias $V_B(x)$
(see Fig.~\ref{potentials})
\begin{subequations}
\begin{align}
V_A(x) &= V_0 \left( x^4 - x^2 \right) \\
V_B(x) &= V_0 \left( x^4 - x^2 + x \right) .
\end{align}
\end{subequations}

The free energy difference for a quantum particle confined by the
potentials $V_A$ and $V_B$ will be computed from the Crooks
fluctuation relation using the time-dependent potential
$V(x,\lambda(t)) = V_A(x) + \lambda (t) V_0 x$, where we assume
$\lambda (t) = t/\tau$ and $\tau$ defines the rate at which the
potential is switched. Note that $\lambda_A=\lambda(0)=0$ and
$\lambda_B=\lambda(\tau)=1$. The bead regularization will be used
because of its better convergence properties compared to the Fourier
regularization.

For application of the non-equilibrium relations, the procedure
consists of drawing $N$ initial values of the bead positions
$\mathbf{x} = \{ x_1, \dots, x_M\}$ and conjugate momenta $\mathbf{p}
= \{ p_1, \dots, p_M\}$ from the canonical probability density
\begin{equation*}
\rho(\mathbf{x},\mathbf{p}) = e^{-\beta H_M(\mathbf x,\mathbf p, 0)}/Z_A
\end{equation*}
and propagating each phase point $(\mathbf x_i,\mathbf p_i)$
($i=1\ldots N$) forward to time $\tau$ under the influence of the
time-dependent potential $V(\mathbf x,\lambda (t))$.

\subsection{Sampling initial conditions}

The sampling of initial conditions for the dynamics was done using a
Monte Carlo procedure.  For systems that have weak quantum character,
i.e., with large $\kappa = m/(\beta \hbar)^2$, the harmonic part
\begin{equation*}
U_h(\mathbf x) = \frac12\kappa M \sum_{n=1}^{M} \left( x_n - x_{n+1} \right)^2
\end{equation*}
of $H_M$ arising from the kinetic energy operator of the quantum
particle, forces the bead positions $x_n$ to be near one another.
This strong harmonic potential $U_h(\mathbf x)$ makes Monte-Carlo
sampling of the canonical distribution inefficient if the sampling is
based only on trial moves generated by uniformly chosen random
displacements of the bead positions $x_n$.  For this reason, it is
helpful to use importance sampling based on the probability density
for a free particle.  The procedure consists of dividing the trial
configurations generated in the Monte-Carlo sampling into two types.
The first type consists of randomly displacing the centroid or
center-of-mass of the ring polymer.  Since the potential $U_h$ is
constant for this type of displacement, only the potential $V(\mathbf
x, \lambda)$ determines the acceptance probability of this type of
move.  The second way of generating trial configurations consists of
drawing independent, non-zero Fourier modes $\{ \tilde x_k \}$ with
$k=1$ to $k=\floor{\frac{M-1}{2}}$ based on the probability density
\begin{subequations}
\begin{equation}
P( \tilde x_k^{(r)}, \tilde x_k^{(i)} ) = \frac{a_k}{\pi} e^{- a_k
\left( \tilde x_k^{(r)2} + \tilde x_k^{(i)2}\right)} ,
\end{equation}
where $a_k = \beta m \omega_k^2$ and $\tilde x_k^{(r)}$ and $\tilde
x_k^{(i)}$ are the real and imaginary parts of $\tilde x_k$,
respectively.  Note that if $M$ is even, then $\tilde x_{M/2}^{(i)} =
0$ and $\tilde x_{M/2}^{(r)}$ is drawn from
\begin{equation}
   P( \tilde x_{M/2}^{(r)} )  
   = \sqrt{\frac{b_k}{\pi}} e^{- b_k \tilde x_{M/2}^{(r)2} } ,
\end{equation}
\end{subequations}
with $b_k = \frac12\beta m\omega_k^2$.  The Fourier modes $\{ \tilde
x_k \}$ are then inverted to form the positions of the beads $\mathbf
x$ for the trial configuration, which is then accepted with
probability $\text{min} \left( 1, e^{-\beta \Delta V} \right)$, where
$\Delta V$ is the difference in the potential between the trial and
original positions (i.e. with no harmonic contribution $U_h$).  Since
this procedure generates trial configurations with the same centroid
as the original configuration, $\Delta V$ is typically small and most
trial configurations are accepted if quantum effects are not too
large.  Once a statistically independent configuration of the ring
polymer has been obtained, conjugate momenta $\mathbf{p}$ may be drawn
from a Maxwell-Boltzmann density to obtain an initial phase point for
the system [cf.~Eq.~\eqref{MB}].

\subsection{Dynamics}

Although the numerical time propagation of the system can be carried
out in a multitude of ways, symplectic integration methods usually
offer superior stability and accuracy.  The general Hamiltonian
\eqref{discreteH} is time-dependent and the dynamics is more
complicated than that of autonomous systems.  Nonetheless, it can be
shown that phase space volume still satisfies Liouville's theorem
under the dynamical flow and that integration schemes of second and
higher order can be derived for the time-dependent potentials in the
same manner as for autonomous systems, as long as the time is updated
only after the momentum-propagation step\cite{Suzuki}. For our test
system, we used a second-order Verlet-like integration scheme that
satisfies this requirement, where first the momenta are propagated for
a half-step $\delta t/2$ using initial forces, then an update of the
positions and the system time is done using the current momenta by a
full step $\delta t$.  The time-dependent force is then computed at
this system time using the updated positions, and finally a final
momentum update of a half-step $\delta t/2$ is carried out.  It is
readily established that a trajectory using this scheme is exact for a
Hamiltonian that differs from $H_M$ by terms of order $\delta t^2$ at
all times.

\subsection{Estimating the free energy difference}

After each phase point has been propagated to time $\tau$, the
non-equilibrium work $w_i = H_M(\mathbf x_i(\tau), \mathbf p_i(\tau),
1) - H(\mathbf x_i(0), \mathbf p_i(0), 0)$ is computed. From a set of
$n$ initial conditions and work values, the free energy difference for
the quantum system can then be computed using the Jarzynski estimator
\begin{equation}
   -\beta \Delta F =
   \log \left( \frac{1}{N} \sum_{i=1}^{N} e^{-\beta w_i} \right)
.\label{expaverage}
\end{equation}
Statistical uncertainties for this estimator may computed using
jackknife\cite{Jackknife} or bootstrap\cite{Bootstrap,NumRecipes}
methods on the sample.

As is clear from Eq.~\eqref{expaverage}, the free energy difference computed
from the exponential average of the work is sensitive to large
fluctuations in the value of the work $w_i$ and may converge slowly if
there are large tails in the work
distribution\cite{JarzynskiConvergence}.  The Crooks fluctuation
relation, on the other hand, is based on finding the value of the work
$W_c$ at which the probability density of the work $P_f(W_c)$ is equal
to the probability density $P_r(-W_c)$ of the negative of the work in
the reverse process in which initial conditions are drawn from a
canonical distribution based on the potential $V_B(\mathbf x)$.
According to the Crooks relation in Eq.~\eqref{CFR}, the free energy
difference is then given by $\Delta F = W_c$.  Calculations of the
free energy difference based on the Crooks fluctuation relation
therefore require constructing the probability densities of the work
in the forward and reverse directions.  The traditional approach of
approximating such densities is to use histograms.  While the use of
histograms is parameter free for systems in which the variable is
confined to discrete values, the more typical situation concerns
probability densities of continuous variables, which requires the
specification of a bin size for representing the density.  Thus, in
contrast to the calculation of the free energy difference using the
Jarzynski fluctuation relation, the use of a histogram approach to
approximating probability densities leads to an undesirable parameter
dependence to the free energy differences computed in the Crooks
fluctuation approach.

In fact it is possible to reconstruct probability densities of
continuous variables without resorting to parameter-dependent
histogram methods\cite{BergHarris07}. The idea is to expand the
empirical cumulative distribution function (ECDF) obtained after
sorting the data in a series of complete orthogonal polynomials.  From
the mathematical properties of the cumulative distribution function,
the number of terms required in the expansion of the ECDF can be
determined without user intervention by application of the Kolmogorov
or Kuiper's test\cite{NumRecipes}.  From this expansion, an analytical
form for the probability density can be obtained by differentiation.
This approach can be applied to the probability densities of the work
for both the forward and reverse processes.  The free energy is then
obtained numerically by finding the value of the work where the
analytical probability densities are equal.

In principle, any orthogonal set of polynomials can be utilized for
the expansion of the ECDF.  In Ref.~(\onlinecite{BergHarris07}), a
Fourier series for the difference between the ECDF and a linear
function was used to construct an analytical approximation to the
ECDF.  Although such an approach successfully produces a fairly smooth
approximation of the ECDF, the use of oscillatory transcendental
functions introduces high frequency oscillations in the smooth
approximation that may lead to systematic errors when finding points
of overlap of densities, particularly when the densities intersect in
regions where there are prolonged tails.  As an alternative, we
consider using a high-order polynomial expansion via the Chebyshev
polynomials.

To illustrate the method, consider a series of $n$ {\it sorted} work
values $\{ w_i \}$ where $w_i \leq w_{i+1}$.  An unbiased estimator of
the distribution function of the work $P(w)$ is the ECDF defined in
the range $[w_1,w_n]$ as
\begin{equation}
   \bar{P}(w) = \frac{i}{n} \qquad \mbox{for $w_i \leq w < w_{i+1}$}.
\label{ecdf}
\end{equation}
To allow an expansion in Chebyshev polynomials which are defined on
the interval $[-1,1]$, the ECDF is mapped to that domain using
\begin{align}
   \bar{w} &= \frac{2w - w_1 - w_n}{w_n - w_1} 
\\
   \bar{F}(\bar{w}) &= \bar{P}(w).
\end{align}
The new ECDF $\bar{F}$ can then be expanded in terms of Chebyshev
polynomials of the first kind $T_n$ to yield the approximation
\begin{eqnarray*}
\bar P(w)\approx P_m(w) = F_m(\bar{w}) = \frac{d_0}{\pi} + \frac{2}{\pi} \sum_{j=1}^{m} \,
d_j \, T_j(\bar{w}) ,
\end{eqnarray*}
where the coefficients $d_j$ are given by
\begin{eqnarray*}
d_j &=& \int_{-1}^{1} \frac{\bar{F}(w) \, T_j(w)}{\sqrt{1-w^2}} \, dw .
\end{eqnarray*}
The analytical approximation to the probability density $p_m(w)$ in
terms of $m$ Chebyshev polynomials is then given by
\begin{eqnarray*}
p_m(w) &=& \frac{4}{\pi(w_n - w_1)} \sum_{j=1}^{m} j d_j U_{j-1}(\bar{w}),
\end{eqnarray*}
where $U_n$ are Chebyshev polynomials of the second
kind\cite{AbramowitzStegun}. The expansion coefficients $d_j$ can be
evaluated analytically using the form of the ECDF in \eqref{ecdf}, and
are given by
\begin{eqnarray*}
d_0 &=&  
\frac1n\sum_{i=1}^n \arccos(\bar w_i)
\\
d_j &=& \frac1n\sum_{i=1}^n\frac1j\sqrt{1-\bar w_i^2}\,U_{j-1}(\bar w_i).
\end{eqnarray*}
In practice, one hopes that the number of polynomials $m$ required in
the expansion of the ECDF is modest so that a smooth approximation is
obtained.  What number of polynomials is appropriate can be estimated
using either the Kolmogorov\cite{BergHarris07,NumRecipes} or Kuiper's
test\cite{NumRecipes}, which determine how likely it is that the
difference between the ECDF $\bar{P}(w)$ and its analytical
approximation $P_m(w)$ is due to random variations.  The tests take
the maximum variation between $\bar{P}$ and $P_m$ over the sampled
points and return a probability $Q_m$ that the difference between the
two cumulative distribution functions is due to chance.  A small value
of $Q_m$ indicates that the difference between the cumulative
distribution functions is significant, so that the quality of the
expansion $P_m$ is insufficient to represent the data.  One therefore
carries out a process of progressively increasing the number of
polynomials $m$ and evaluating $P_m$ as well as $Q_m$ until the value
of $Q_m$ is larger than some threshold, say $Q_{c} = 0.5$.

Once analytical approximations to the probability densities of the
work in the forward and reverse directions have been obtained, the
intersection point $W_c$ is readily evaluated by numerically searching
for a solution of $P_f(W_c)=P_r(-W_c)$ using the Brent
method\cite{NumRecipes}.  The Brent method requires that the solution
be bracketed on an interval $(w_{min}, w_{max})$.  Since the free
energy difference between ensembles are single-valued, there is only
one point of intersection of the probability densities in the forward
and reverse directions.  The interval end-points $w_{min}$ and
$w_{max}$ can therefore be taken to be near the extremal values of the
$w_i$ values found in the simulations.

Finally, statistical uncertainties can also be computed for the free
energy difference by repeating the calculation of the intersection
point for a series of jackknifed samples of the data and using the
variance of the free energy differences over the jackknife samples.

\begin{figure}[t]
\includegraphics[width=0.9\columnwidth]{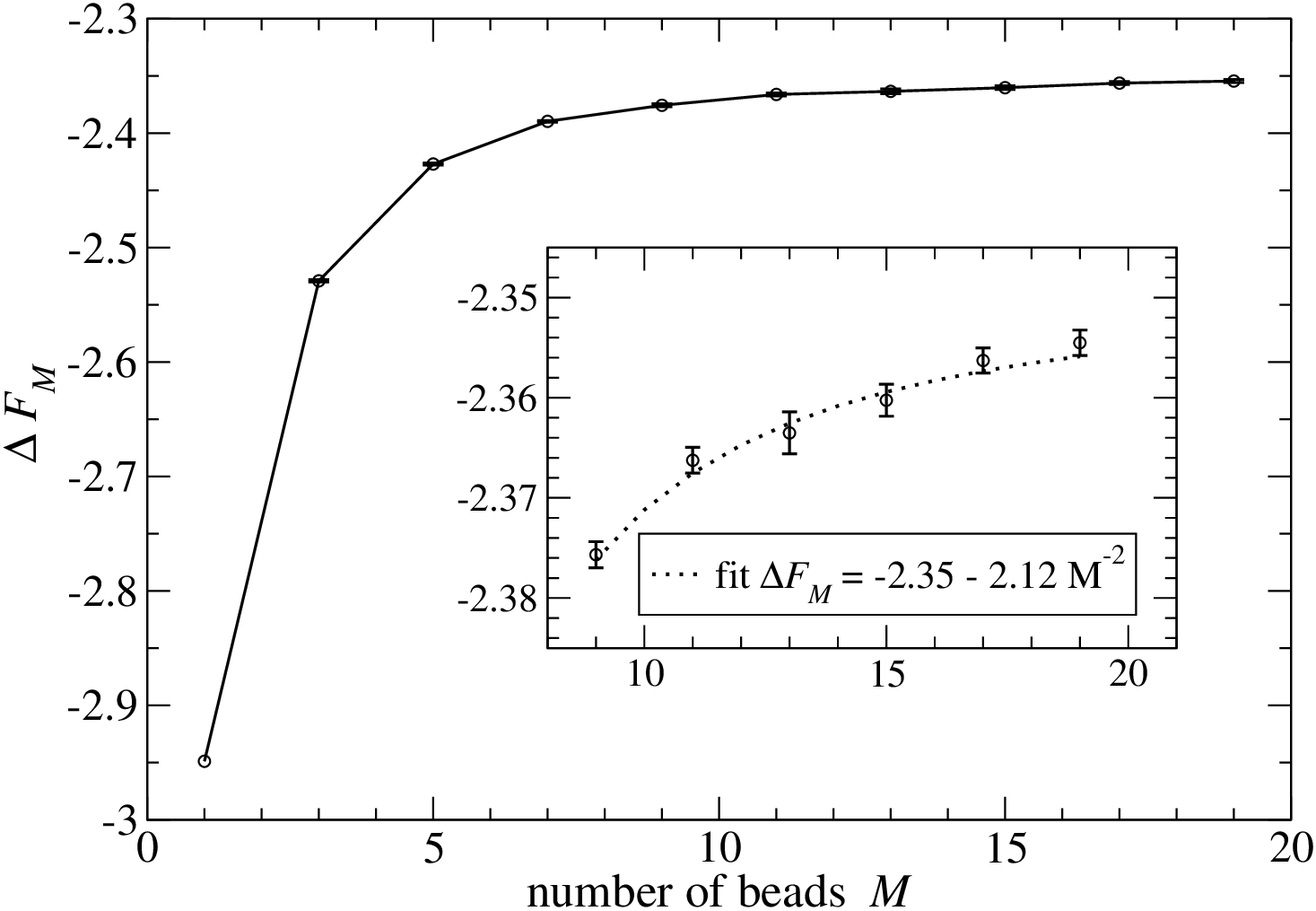}
\caption{The difference in free energy for the quartic system as a
         function of the level of discretization of the path integral
         for the fast-switching non-equilibrium process
         ($\tau=0.5$). The solid line is drawn as a guide to the eyes,
         while the dotted line in the inset is a fit to $a+bM^{-2}$,
         as predicted by the theory, which agrees within errorbars.}
\label{freeEresult}
\end{figure}

\subsection{Simulation results}

To illustrate this approach, numerical tests were carried out to
calculate the quantum free energy difference between systems with
$\lambda = 0$ (symmetric quartic double-well potential) and $\lambda =
1$ (biased quartic potential) with parameters $\beta = 1$, $\hbar =
1$, $m=1$ and $V_0 = 5$, while the mass $\mu'$ per bead was also set
to $1$.  The time step for the propagation was $\delta t = 0.001$ such
that the fluctuations in the total energy at constant $\lambda = 0$
relative to the fluctuations in potential energy were less than $1\%$.
The calculations were carried out at two different switching rates,
$\tau = 0.5$ (fast switching) and $\tau = 100$ (slow switching).  

In Fig.~\ref{freeEresult}, the free energy difference $\Delta F_M$
calculated using the Crooks fluctuation relation for fast-switching
process is shown as a function of the number of beads $M$ in the ring
polymer regularization of the path integral.  The work value $W_c$ at
which $P_f(W_c) = P_r(-W_c)$ was based on the analytical approximation
of the empirical cumulative distribution functions formed out of $1
\times 10^7$ independent realizations of the non-equilibrium process
in the forward and reverse directions.  The inset graph in this figure
clearly suggests that $\Delta F_M$ converges as $1/M^2$, to a final
value of $\Delta F=-2.35$. This value agrees with the quantum free
energy difference found by evaluating the partition sum using the
numerically determined eigenvalues of the Hamiltonians $\hat H(0)$ and
$\hat H(1)$.  The free energy difference is not just due to the
difference in zero-point energy, which is $-2.53$, but to the higher
energy levels as well.  Note also that the quantum contributions to
the free energy difference lead to a quantum free energy difference
that is roughly $25\%$ higher than the classical value of $-2.95$.

\begin{figure}[t]
\includegraphics[width=0.9\columnwidth]{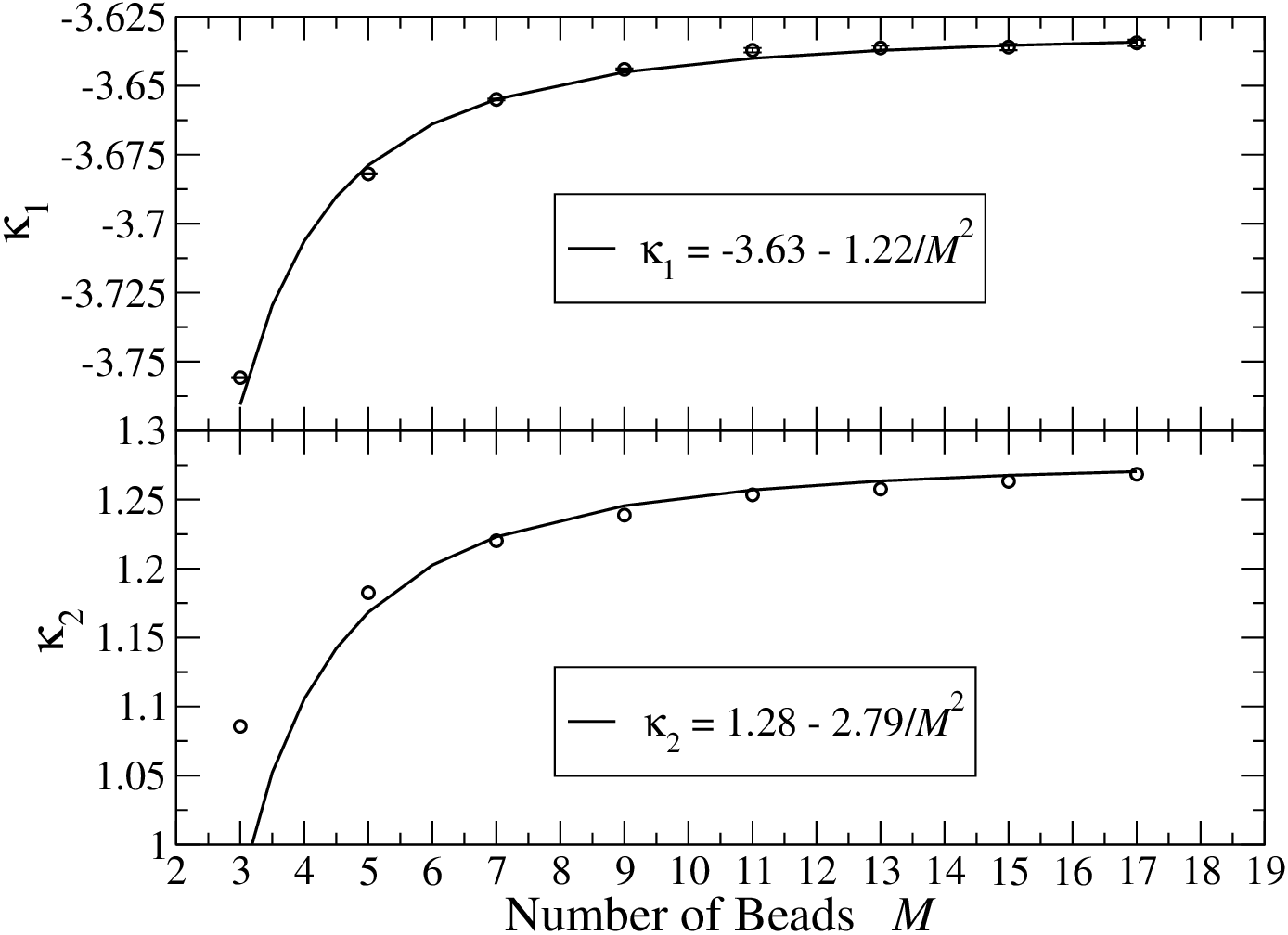}
\caption{The first and second cumulants of the work distribution for
         the reverse process for the quartic system as a function of
         the level of discretization of the path integral for the
         fast-switching non-equilibrium process ($\tau=0.5$). The
         solid lines are a fit to $a+bM^{-2}$.}
\label{kappa12result}
\end{figure}

In Fig.~\ref{kappa12result}, the first and second cumulants of the
work done (i.e., the average and the variance of the work) in the fast
reverse process are plotted against $M$.  As the fits in the
figures show, the asymptotic convergence of both cumulants is
consistent with a $\order{M^{-2}}$ behavior. The results presented in
this figure suggest, in general, that the work distributions converge
in the infinite $M$ limit as $\order{M^{-2}}$, which can be
rationalized from the consideration of a harmonic oscillator system
(see the companion paper\cite{subsequent}).

\begin{figure}[t]
\includegraphics[width=0.9\columnwidth]{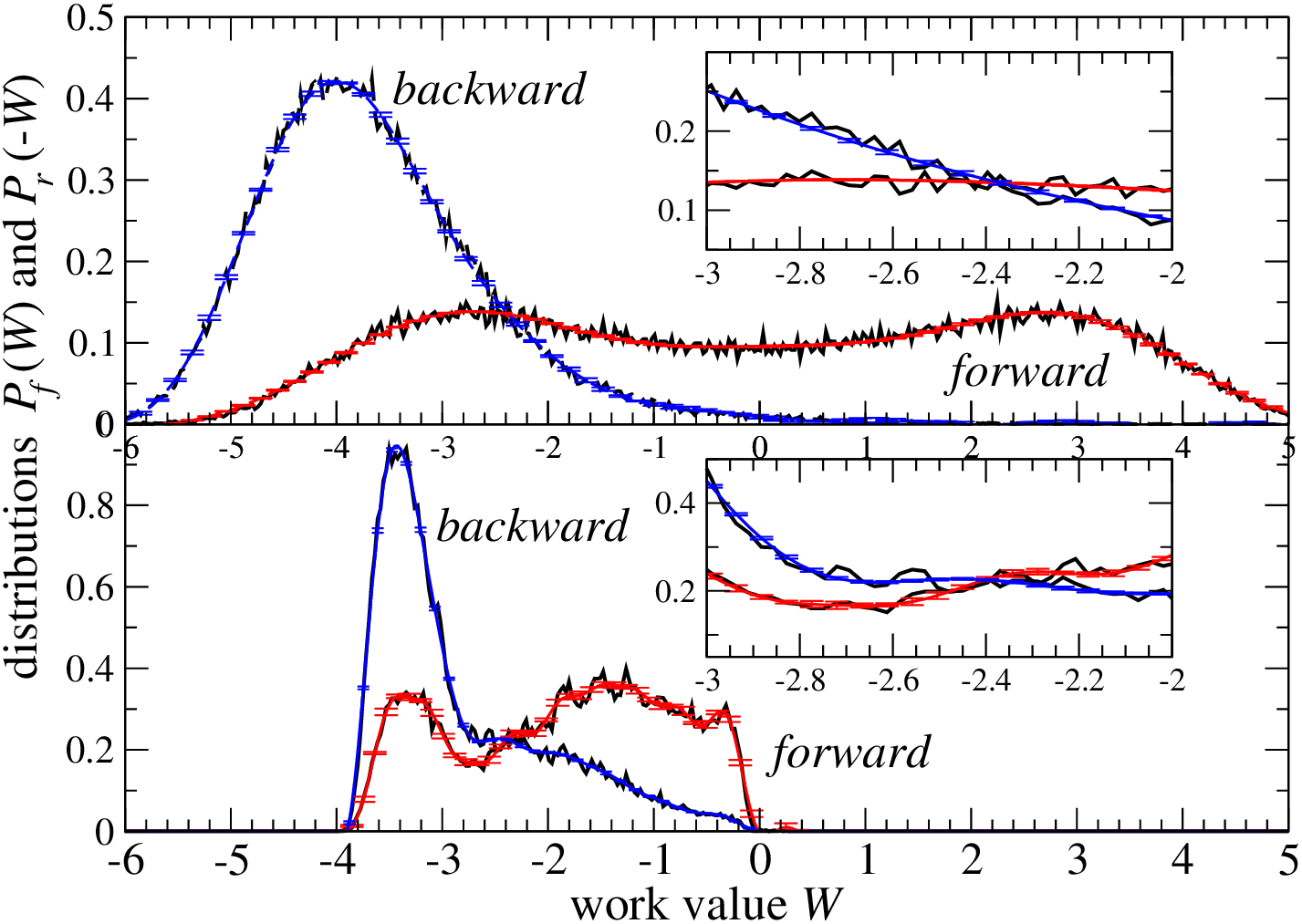}
\caption{(Color online) The forward and backward work distributions in
  histogram and analytical expanded form based on $1 \times 10^5$
  realizations of the non-equilibrium process for the quartic system,
  using $M=19$.  The top panel corresponds to results for the
  fast-switching process with $\tau = 0.5$ while the lower panel
  contains the results for a slow-switching process with $\tau =
  100$. Noisy curves correspond to the histogram form and the smooth
  curves to the analytical expanded form of the work
  distributions. The insets show the detail in the vicinity of the
  crossing point.}
\label{crooksResults}
\end{figure}

The probability densities of the work and negative work in the forward
and reverse directions are shown in Fig.~\ref{crooksResults} as
histograms and expanded analytical forms.  The
convergence properties of the analytical expansions
of the empirical cumulative distribution function were determined by
the asymptotic Kuiper's test with a threshold value $Q_c$ of $0.5$.
Typically, $m=13$ terms were required to reach convergence with this
choice of $Q_c$.  The probability densities shown in
Fig.~\ref{crooksResults} were estimated from $1 \times 10^5$ values of
the work and negative work for the non-equilibrium switching process.
The data shown are for a $M=9$ bead discretization of the path
integral.  Note that even though the shape of the probability
densities is sensitive to the switching rate, the intersection point
of the forward and reverse densities is independent of the rate and
equal to the free energy difference.  It is apparent from the detail
of the work densities near their point of intersection shown in the
inset of the panels that it is difficult to arrive at an estimate of
the free energy difference based on noisy histograms of the work.  In
contrast, the analytical forms of the probability densities lead to
smooth curves and unambiguous points of intersection.

\section{Conclusions}
\label{conclusions}

Non-equilibrium methods for the calculation of free energy differences
in quantum systems in the context of the path integral representation
of the canonical partition function have been presented. Instead of
using the real quantum dynamics of the system, the path integral
representation allows a fictitious path to be defined for which the
Jarzynski and Crooks relations are valid. By evolving the ring polymer
in the path integral representation under fictitious dynamics, the
difficulties associated with the complexity of the full evolution of a
quantum system are avoided.  {}From a computational perspective,
avoiding true quantum dynamics is a great advantage, although the
actual efficiency as compared to other methods to compute free energy
differences will depend greatly on implementation details and on using
additional techniques such as importance sampling.

{}From a formal point of view, the path integral approach exploits the
well-known isomorphism between a quantum system and a classical system
of higher dimensionality described by a field theory.  While the
dynamical evolution of the classical field is well-defined, the
isomorphic classical system exhibits the typical divergent behavior of
classical field theories on short length scales,
due to the infinite number of degrees of freedom.
It was demonstrated that the
equations of motion and field variables can be regularized using
either a wave-vector cut-off of the Fourier modes of the fields $x(u)$
and $p(u)$, or a real space discretization of the ring polymer
representing the quantum particle, yielding a finite number $M$ of
degrees of freedom.

A general numerical procedure for calculating the free energy
difference between non-trivial quantum systems was elaborated using a
particle confined in a quartic potential as a test model.  General
issues pertaining to sampling the initial phase points, performing the
dynamics of non-autonomous systems, and estimating free energy
differences using the exponential average of the work (Jarzynski
method) and the crossing method (Crooks approach) were discussed.  A
parameter-free method for calculating the free energy difference using
the crossing of the forward and reverse work distributions was
introduced.  The parameter-free method is based on expanding the empirical cumulative
distribution function in orthogonal Chebyshev polynomials and is
controlled by a rigorous statistical convergence test. From these expansions,
analytical functions of the estimated forward and reverse work
distributions are then obtained from which precise values of the
crossing point and hence the free energy can be extracted.

The expansion procedure utilized here is quite general, and can be
used to construct analytical estimates of any probability density of a
single continuous variable.  The approach should be particularly
useful whenever the specific value of a probability density if
desired, such as in the calculation of the value of the radial
distribution function at contact in a system of hard spheres.  The
method could also be of use in construction of the free energy as a
function of a reaction coordinate or other parameter, where the
dispensing of parameter-dependent histograms is desired.

The numerical results indicated that for the quartic potential the
work distribution converges as the regularization parameter $M$ gets
large, and hence so does the free energy estimate obtained from the
crossing method.  Although the general convergence of the free energy
and work distribution utilized in the crossing method must be
demonstrated on a case-by-case basis, the following paper shows that
for the harmonic oscillator the work distribution and free energy
converges rigorously but that the nature of the convergence depends on
the regularization used\cite{subsequent}.  In particular, it is shown
that the bead regularization converges faster than the Fourier
regularization.

While the focus here has been on deterministic methods with
Hamiltonian dynamics in non-equilibrium statistical mechanics, it is
worth mentioning that an alternative to the use of Hamiltonian
dynamics is to evaluate the path integral via the path integral Monte
Carlo (PIMC) method. Both the Jarzynski and Crooks relations hold in
this context provided that the process is Markovian and time
reversible\cite{Jarzynski2,Crooks1,Crooks2}. Furthermore, for a given
regularization at finite $M$, one can in principle also use
thermostatted deterministic dynamics, for which the Jarzynski and
Crooks relations also hold\cite{Dellago}. The convergence properties
of these alternative non-equilibrium processes as $M\to\infty$ will be
assessed in future work.

Finally, it should also be noted that the dynamics generated both
through deterministic and stochastic evolution is completely
artificial and generally has little to do with the real time quantum
dynamics of the system (except when $\hbar\beta\to0$).  Still,
the computation of the free energy difference through the dynamical
non-equilibrium procedure described above yields exact quantum results
in the limit $M\to\infty$. Moreover, the use of either deterministic
or stochastic evolution bypasses the problem of computing the real
time quantum dynamics and illustrates the power of path integral
methods in practical applications.

\acknowledgments

The authors would like to acknowledge support by grants from the
Natural Sciences and Engineering Research Council of Canada NSERC.

\end{document}